\documentclass[usenatbib]{mn2e}
\usepackage{times}
\usepackage{epsf}
\usepackage{natbib}
\usepackage{fixltx2e}
\newcommand{\etal}{{et al}\/.}

\begin{document}
\title[X-rays from the EELR of 3C\,171]{X-ray emission from the extended emission-line region of the
  powerful radio galaxy 3C\,171}
\author[M.J. Hardcastle \etal]{M.J.\ Hardcastle$^1$\thanks{E-mail: m.j.hardcastle@herts.ac.uk}, F. Massaro$^2$ and D.E.
  Harris$^2$\\
$^1$ School of Physics, Astronomy \& Mathematics, University of
  Hertfordshire, College Lane, Hatfield AL10 9AB, UK\\
$^2$ Harvard-Smithsonian Center for Astrophysics, 60 Garden Street, Cambridge, MA~02138, USA}
\maketitle
\begin{abstract}
We present {\it Chandra} X-ray observations of the powerful radio
galaxy 3C\,171, which reveal an extended region of X-ray emission
spatially associated with the well-known 10-kpc scale optical
emission-line region around the radio jets. We argue that the X-ray
emission comes from collisionally ionized material, originally cold gas
that has been shock-heated by the passage of the radio jet, rather
than being photoionized by nuclear radiation. This hot plasma
is also responsible for the depolarization at low frequencies
of the radio emission from the jet and hotspots, which allows us to
estimate the magnetic field strength in the external medium. We show
that it is likely that both the cold emission-line gas and the hot
plasma in which it is embedded are being driven out of the host galaxy
of 3C\,171 at supersonic speeds. A significant fraction of the total
energy budget of the central AGN must have been expended in driving
this massive outflow. We argue that 3C\,171, with its unusual radio
morphology and the strong relation between the jet and large amounts
of outflowing material, is a member of a class of radio galaxies in
which there is strong interaction between the radio jets and cold
material in the host galaxy; such objects may have been very much more
common in the early universe.
\end{abstract}

\section{Introduction}

In powerful radio galaxies, extended optical emission-line regions
(EELR) are often found to be aligned with the axis defined by the
extended radio emission (e.g., \citealt{mvbs87}; \citealt{m93};
\citealt*{mbs96}). A long-standing question
\citep[e.g.,][]{bh89,tmrd98} is whether these EELR are ionized by
photons from the nucleus or by shocks driven by the jets. Shocks are
part of the standard model for these powerful objects
\citep[e.g.,][]{bc89} but direct observational evidence for them has
been somewhat elusive, and to date has mostly come from the X-ray band
(e.g., \citealt{kvfj03}; \citealt{omwk06}; \citealt*{wsy06}). However,
in the
most powerful radio sources, normally found at high redshift, there is
often evidence from one or more of the line kinematics, the line
ratios or the relationship with the radio morphology that the optical
emission lines are either ionized or at least strongly affected by
jet-driven shocks, which requires a direct interaction between the
jets and the {\it cold} ($T \la 10^4$ K) phase of the IGM of the host
galaxy (e.g., \citealt*{stb02}; \citealt{nldg08}). Photoionization
presumably still operates, as it must in any source with a radiatively
efficient accretion flow, but the interaction between jets and cold
material dominates the energetics of the EELR. Understanding such
interactions is important in models in which accretion onto the
central black hole affects the properties of galaxies hosting powerful
AGN (`feedback'), while physical conditions in and kinematics of the
EELR in these systems may give us important information about the
energetics of the radio source as a whole. Unfortunately, in nearby,
comparatively low-luminosity radio galaxies, which have the advantage
that they can be studied with high spatial resolution, EELR are
generally small, centrally peaked and can adequately be described by
central-illumination models; they are plausibly just powerful versions
of the ionization cones seen in local radio-quiet AGN such as Seyfert
galaxies \citep[e.g.,][]{bh89}. Thus the study of EELR in nearby
radio-loud sources does not in general shed any light on jet-gas
interactions.

There is evidence, however, that a small minority of low-redshift radio
galaxies are also more adequately described by a shock-ionization
model. One good candidate for a shock-ionization source at low
redshift is 3C\,171, an unusual $z=0.2384$ radio galaxy.
Low-resolution radio imaging \citep*{hvm84,b96} shows that in its inner
regions it is similar to a normal FRII, but low-surface-brightness
plumes instead of normal lobes extend north and south from the
hotspots. High-resolution images show knotty jets connecting the radio
core with the hotspots \citep{hapr97}. 3C\,171 is associated with a
prominent EELR, elongated along the jet axis, and the emission-line
gas is brightest near the hotspots \citep{hvm84}, suggesting that the
radio-emitting plasma is responsible for powering the emission-line
regions. \cite{catr98} made William Herschel Telescope observations of
the emission-line regions and showed that their ionization states are
most consistent with a shock-ionization model, with the disturbed
kinematics and high-velocity line splitting being indicative of direct
driving of motions in the line-emitting gas by the jets. The velocity
centroids of the lines observed are blueshifted at the level of a few
hundred km s$^{-1}$ relative to the systemic velocity in the east and
similarly redshifted in the west, consistent with a jet-driven outflow
(note that the two-sided appearance of the radio jets implies, given
the normal assumption of relativistic jet speeds, that the jets are
not far from the plane of the sky, in which case these radial
velocities may substantially underestimate the outflow speed of the
cold gas). The results of \citeauthor{catr98} were confirmed by later
studies with integral-field spectroscopy and with the {\it Hubble
  Space Telescope} ({\it HST}) \citep{st03,totw05}. In terms of its
apparent strong interactions between jet and cold IGM, 3C\,171 may be
a low-redshift analog of EELR around radio galaxies at high redshift,
where cold gas is expected to be more abundant in radio source host
galaxies.

3C\,171 is also the only radio galaxy to date in which physical
conditions in the EELR have been probed with detailed radio
depolarization studies. \citet{hvm84} pointed out the
depolarization of the radio source near the hotspots, while
\citet{hapr97} showed that the depolarization was spatially very
closely connected with the EELR. More recently \cite{h03} (hereafter
H03) used high-frequency radio polarimetry to measure the
depolarization and Faraday rotation in the source as a function of
position. Radio depolarization of the type seen in 3C\,171 comes about
as a result of unresolved structure in the Faraday rotation due to
magnetoionic material in front of the source. As Faraday rotation
depends on the electron density and magnetic field strength integrated
along the line of sight, measurements of depolarization provide a
measurement of the product of electron density and magnetic field
strength in the external medium \citep{b66} which is essentially
independent of the other physical conditions in that medium, such as
temperature, so long as it is ionized. The key result of H03 was that
the line-emitting material of the EELR, though spatially coincident
with the observed depolarization, cannot itself be responsible for the
depolarization (given the physical parameters deduced from the
observations and the known physical parameters of line-emitting
clouds, the clouds have far too low a covering factor). Instead the
EELR-emitting clouds must be embedded in a more diffuse, and
presumably hotter, medium whose parameters were constrained by the
radio data.

X-ray studies of 3C\,171 allow a search for this diffuse phase of the
external medium of the source. At the time that H03 was writing the
only imaging X-ray data available were a {\it ROSAT} HRI observation
that had previously been shown \citep{hw99} to give no detection of
extended emission. Based on this and on assumed values for the
characteristic magnetic field strength in the external medium, H03
placed some limits on the properties of the hot phase which suggested
that an X-ray detection with deeper observations was unlikely.
However, more recently, a short {\it Chandra} observation taken on
2007 Dec 22 (obsid 9304), with an effective sensitivity a factor of a
few deeper than that of the {\it ROSAT} observations and a much higher
spatial resolution, (Massaro \etal\ in prep) showed extended X-rays
that clearly correlated very well with the observed region of radio
depolarization, implying that the depolarizing medium could indeed be
imaged directly.

The results of the {\it Chandra} snapshot motivated us to make the
deep {\it Chandra} observation that is the subject of the present
paper. In this paper we first describe the deep X-ray observation
(Section 2) and
the principal imaging and spectroscopic results (Section 3). In Section 4 we first
argue that the extended X-ray emission is definitely thermal in origin
and associated with the depolarizing medium and then go on to discuss
the implications of this model for the dynamics and energetics of the
source. Our conclusions are summarized in Section 5.

Throughout the paper we assume a cosmology with with $H_0 = 70$ km
s$^{-1}$ Mpc$^{-1}$, $\Omega_{\rm m} = 0.3$ and $\Omega_\Lambda =
0.7$. This gives a luminosity distance to 3C\,171, at $z = 0.2384$, of
1194 Mpc and an angular scale of 3.77 kpc arcsec$^{-1}$.

\section{Observations}

We observed 3C\,171 with the ACIS-S array on {\it Chandra} on 2009 Jan
08 for a livetime of 59461.4 s (obsid 10303). The observations were
taken in VFAINT mode to allow the best possible background rejection.
Processing was carried out in the standard manner using {\sc ciao}
4.1. In addition to applying the VFAINT background rejection
corrections, we removed the standard 0.5 pixel position randomization
during processing. There were no times of high background (flares)
during the observation and so we use all of the data in the analysis
that follows.

3C\,171 was observed with {\it XMM-Newton} on 2009 Apr 07 as part of
an unrelated program of observations of the nuclei of
intermediate-resolution 3CRR sources \citep*{hec09} for a livetime
before filtering of 18320.8 s (pn) and 23044.6 s (MOS), giving a total
sensitivity in principle very similar to that of the {\it Chandra}
observation. Unfortunately this observation was badly affected by high
particle background -- the count rate in the pn observation was
typically an order of magnitude above the expected background level
between 10 and 15 keV, and the MOS cameras were almost as badly
affected. This effectively removes our ability to use the {\it XMM}
data to study the large-scale environment of the source, and so in
this paper we use them only for comparison with the small-scale
spectroscopy from {\it Chandra}. For these purposes high background is
not so problematic, and so we lightly filtered the pn data to remove
the worst of the flaring, giving an effective observation time of
13101.8 s, and left the MOS data unfiltered.

For comparison with the X-ray we use the radio data described by
H03, the 1.4-GHz image from the 3CRR Atlas\footnote{See
  http://www.jb.man.ac.uk/atlas/ .}, the ({\it HST}) data in the [OIII] and [OII] lines described
by \citet{totw05}, and the ground-based H$\alpha$ image of
\cite{tvmb00}, the last two kindly provided by Clive Tadhunter. For our
analysis we aligned the peak of the nuclear component in the {\it
  Chandra} data (see below) with the position of the radio core. This
required a shift of 0.22 arcsec in a roughly E-W direction, which is
well within the known uncertainties on {\it Chandra}'s absolute
astrometric accuracy\footnote{See
  http://asc.harvard.edu/cal/ASPECT/celmon/ .}. As we have no other
compact features in common between the radio and X-ray maps, we retain
the roll angle of the satellite from the default astrometry. The {\it
  HST} images were also aligned with the radio core
at their brightest point; as this region is resolved with {\it HST}
the alignment is probably only good to $\pm 0.2$ arcsec.

\section{Results}

\subsection{Imaging}

\begin{figure*}
\epsfxsize 16.5cm
\epsfbox{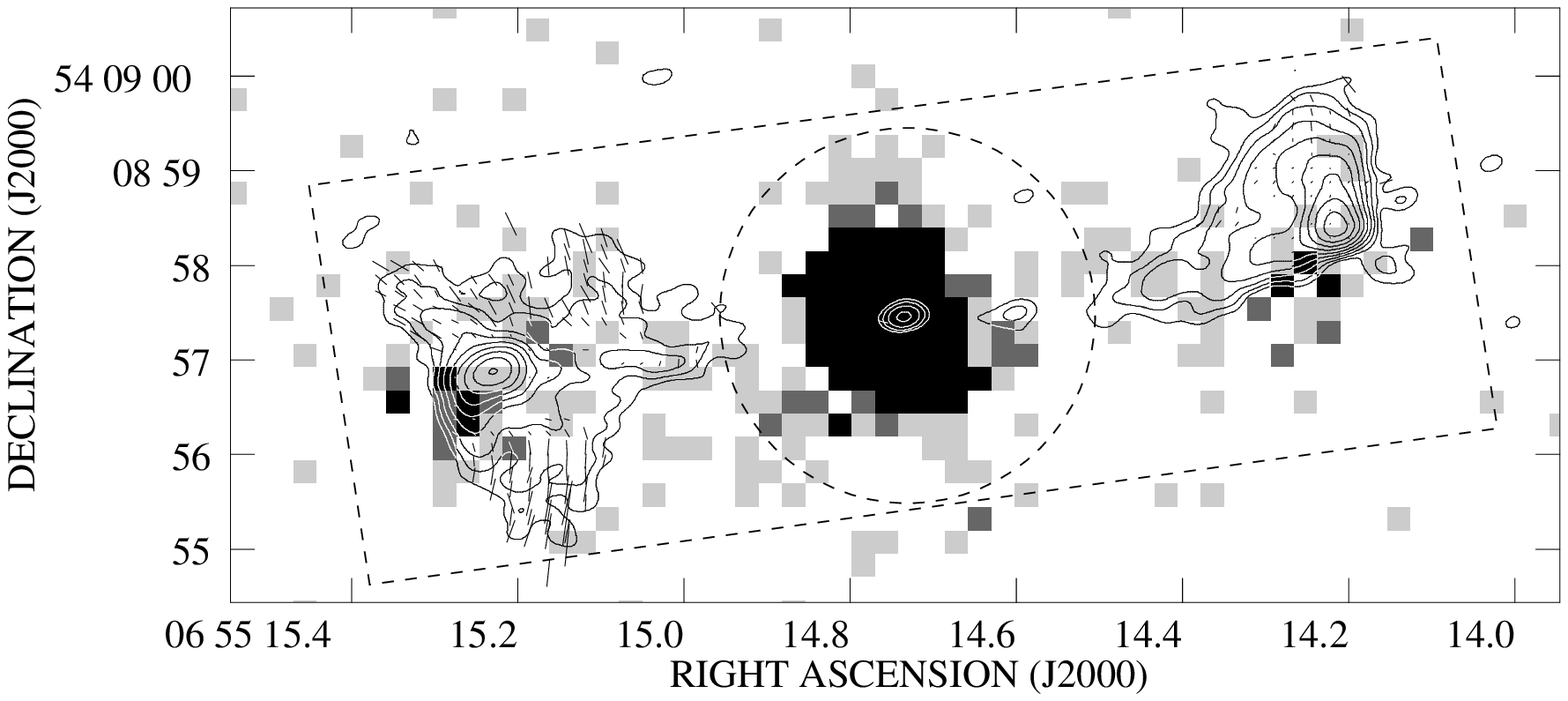}
\caption{Unsmoothed grayscale of {\it Chandra} counts in the energy
  range 0.5-5.0 keV, binned with pixels of 0.5 the nominal {\it
    Chandra} pixel size. Overlaid radio contours are from the 8.1-GHz
  map of H03 with resolution $0.29 \times 0.20$ arcsec. The vectors
  show fractional polarization and polarization direction; they are
  rotated through 90 degrees from the E-vector so that in the absence
  of Faraday rotation they would give an estimate of the magnetic
  field direction in the radio source. Where vectors are not plotted
  no polarization has been detected at 8.1 GHz. Overplotted (dashed
  lines) is the region used for extraction of the X-ray spectrum of
  the extended emission; the region inside the inner circle is
  excluded.}
\label{small-scale}
\end{figure*}

\begin{figure*}
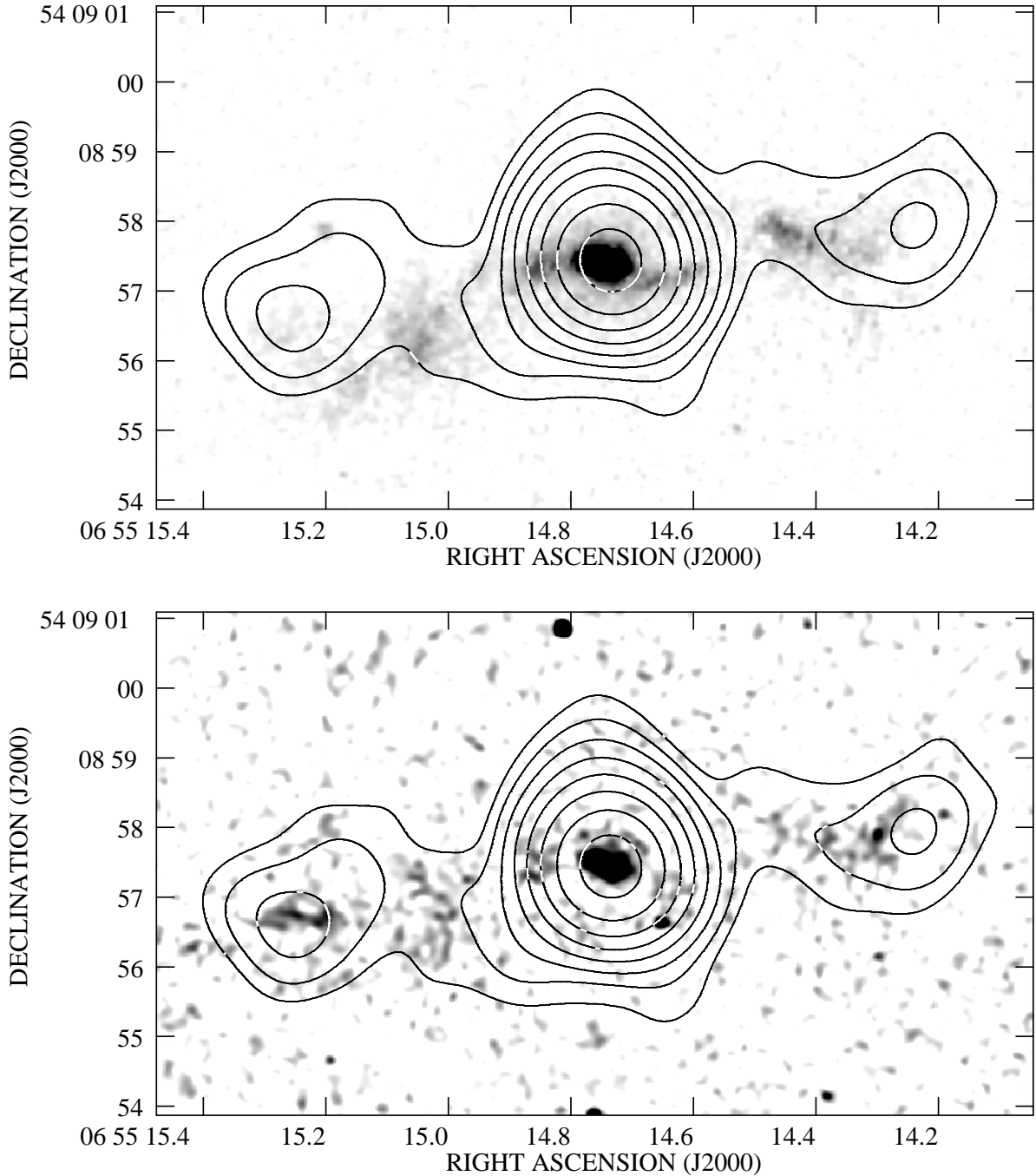

\epsfxsize 16.5cm
\epsfbox{3C171-CH-OIII.PS}
\epsfxsize 16.5cm
\epsfbox{3C171-CH-OII.PS}
\caption{Contours of the 0.5-5.0 keV {\it Chandra} data of Fig.\ \ref{small-scale}, smoothed with
a Gaussian with FWHM 1.0 arcsec, are overlaid on (top) the unsmoothed
{\it HST} [OIII] image and (bottom) the {\it HST} [OII] image smoothed
with a Gaussian of FWHM 0.1 arcsec to improve signal-to-noise. The
lowest contour is at 0.2 counts per $0.246$-arcsec pixel;
this is a factor 2 above the $3\sigma$ level, defined by the method of
\citet{h00}, relative to the off-source background. Contours increase
by a factor 2 at each step.}
\label{emissionline}
\end{figure*}

On small scales, {\it Chandra} resolves the emission from 3C\,171 into
a bright compact nuclear component, possibly slightly extended in a
N-S direction, and low-surface-brightness emission roughly coincident
with the main E-W axis of the radio jet (Fig.\ \ref{small-scale}). In
a 2.5-pixel source circle, with background taken in a concentric
annulus between 2.5 and 4 pixels, there are 1070 0.5-5.0 keV counts in
the core, while the extended emission aligned along the radio jet,
taking a rectangular region with two identical adjacent background
regions and masking out everything within 4 pixels of the nucleus (as
shown in Fig.\ \ref{small-scale}), contains $123 \pm 12$ counts. The
brightest parts of the extended X-ray emission are found near the
radio hotspots, but the brightness peak is offset from the radio in
both cases (0.5-1 arcsec to the S of the W hotspot, and a similar
distance to the E and SE of the E hotspot). In 2.5-pixel source
circles these brightness peaks contain $46 \pm 7$ counts (E peak) and
$32 \pm 6$ counts (W peak) and thus account for the majority of the
X-ray emission from the source.

It is immediately apparent from the polarization vectors in
Fig.\ \ref{small-scale} that there is a good correlation between X-ray
emission and depolarization (seen as unpolarized regions in this
low-frequency radio map): there is no depolarized region that does not
have coincident X-ray emission and only one part of the radio source
coincident with X-ray emission (to the N of the W hotspot) that shows
significant radio polarization. The X-ray emission also broadly
follows the region delineated by the optical emission-line region
(Fig.\ \ref{emissionline}). The clearest differences are near the E
hotspot, where the [OIII] emission runs south of the X-rays (although
the [OII] data show a very similar structure to the X-ray emission)
and at the W hotspot, where the peak brightness in the X-rays has no
corresponding optical emission. The X-rays also appear very similarly
distributed to the H$\alpha$ emission seen in the ground-based imaging
of \citet{tvmb00}. There is clearly an intimate connection
between the X-ray emission, the radio depolarization, and the
optically emitting gas.

\begin{figure}
\epsfxsize 8.5cm
\epsfbox{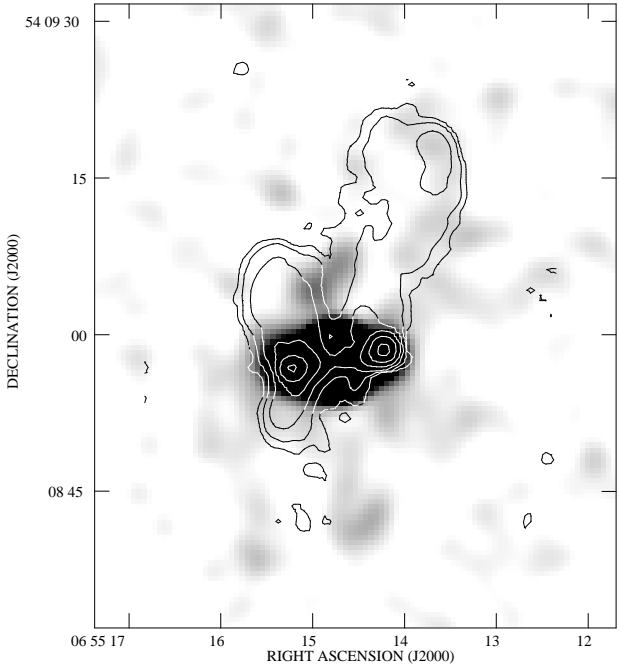}
\caption{{\it Chandra} counts in the energy range 0.5-5.0 keV, binned
  to the nominal pixel size and smoothed with a Gaussian of FWHM 3.0
  arcsec. The burnt-out central region roughly corresponds to the
  region seen in Figs \ref{small-scale} and \ref{emissionline}. Overlaid radio contours are from the 1.4-GHz VLA map from
  the 3CRR Atlas, with a resolution of 1.3 arcsec, and are at $0.5
  \times (1,4,16\dots)$ mJy beam$^{-1}$.}
\label{largescale}
\end{figure}

\begin{figure}
\epsfxsize 8.5cm
\epsfbox{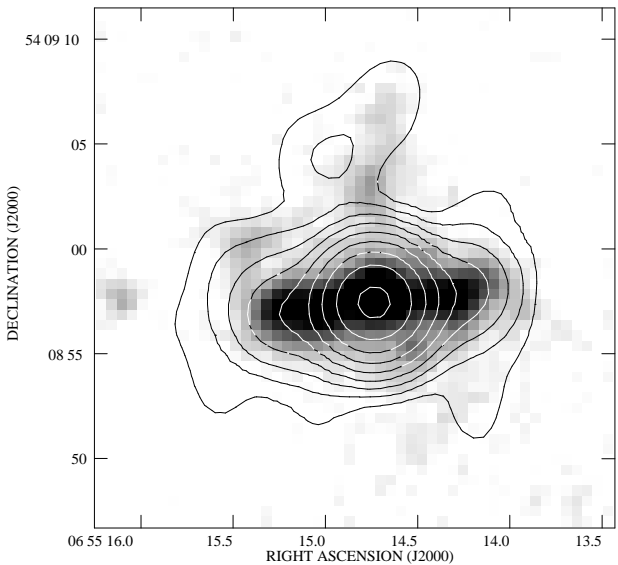}
\caption{H$\alpha$ image of 3C\,171 from \citet{tvmb00} with
  superposed contours from the smoothed X-ray image of
  Fig.\ \ref{largescale}. The lowest contour is at the $3\sigma$ level
  as determined using the method of \citet{h00}, corresponding to
  0.082 counts per $0.492$-arcsec pixel. Contours increase by a factor 2 at
each step.}
\label{halpha}
\end{figure}

On larger scales, there is little evidence of any extended X-ray
emission associated with 3C\,171 (Fig.\ \ref{largescale}). The most
obvious feature extends about 10 arcsec N of the nucleus, and is
roughly coincident with a similar linear H$\alpha$ feature seen by
\citet{tvmb00} (Fig.\ \ref{halpha}). This extended feature contains
$17 \pm 4$ 0.5-5.0 keV counts. Weak extended emission is coincident
with the W radio lobe (Fig.\ \ref{largescale}) but does not appear to
trace the radio structure. At most a few tens of counts in the 0.5-5.0
keV energy band can be associated with all the emission on scales
larger than that of the jet.

\subsection{Spectroscopy}
\label{spectroscopy}

Spectral fitting was carried out using XSPEC 12. Spectra were grouped
to have a minimum of 20 counts per bin after background subtraction.
Spectral fitting to {\it Chandra} data was done in the 0.4-7.0 keV
energy range in which {\it Chandra} is well calibrated, while for {\it
  XMM-Newton} the 0.3-8.0 keV energy range was used. All fits included
Galactic absorption with an assumed $N_{\rm H} = 6.6 \times 10^{20}$
cm$^{-2}$. Errors quoted are $1\sigma$ for 1 degree of freedom
($\Delta\chi^2 = 1.0$).

We began by considering the {\it Chandra} spectrum of the bright core,
which was previously discussed by \citet{hec09}. The core has a count rate of
0.076 photons per 3.2 s frame, so pileup is expected to be negligible. In the
2.5-pixel source region discussed above, the data were well fitted
($\chi^2 = 50.1/59$) using the standard model for narrow-line radio
galaxies, consisting of a power law with only Galactic absorption
together with another power-law component with much higher intrinsic
absorption. The unabsorbed component is required --- fitting with an
absorbed power law only gives $\chi^2 = 76/61$ --- but its photon
index is not constrained and we fix it to the value of 2.0 adopted by
\citet{hec09}. Other forms for this component, such as soft
thermal emission, cannot be ruled out. The properties of the absorbed
component are well constrained; the absorbing column is $(8.8 \pm 1.0)
\times 10^{22}$ cm$^{-2}$ and the photon index is
$1.76_{-0.10}^{+0.19}$. The implied absorption-corrected rest-frame
2-10 keV luminosity of the absorbed component is $1.2 \times 10^{44}$
erg s$^{-1}$. Iron features are often seen in the spectra of
narrow-line FRIIs, but none is present in 3C\,171.

The spectrum of the extended emission, using the region discussed
above, could be fitted either with an APEC model (abundance fixed to
0.35 solar, a typical value for the intra-group medium; $\chi^2 =
5.9/4$, $kT = 1.5 \pm 0.2$ keV) or a power law ($\chi^2 = 3.3/4$,
$\Gamma = 2.7 \pm 0.2$). The presence of some emission at high
energies, and the absence of any peak in the emission around 1 keV,
were the main factors favouring the power-law model. Better fits with
APEC models could be obtained by reducing the abundance, so we do not
regard the difference in $\chi^2$ here as in any way conclusive; with
spectra of this signal-to-noise it is not possible to distinguish
between power-law and thermal models in complete generality. Nor can
we distinguish with these data between single-phase and multiphase, or
between photoionized and collisionally ionized gas; there is no simple
model within XSPEC that would adequately represent a photoionized
plasma. For the thermal model, the (absorbed) 2-10 keV flux is $1.8
\times 10^{-15}$ erg cm$^{-2}$ s$^{-1}$, the (absorbed) flux in the
0.1-2.4 keV {\it ROSAT} band used by H03 is $7.2 \times 10^{-15}$ erg
cm$^{-2}$ s$^{-1}$ (consistent with the H03 upper limit; the flux is
higher than in the hard X-ray band due to the soft spectrum of the
extended emission), and the rest-frame 2-10 keV unabsorbed luminosity
is $4.7 \times 10^{41}$ erg s$^{-1}$.

We investigated whether there was any difference in hardness ratio
between the hotspot regions and the rest of the extended emission. To
do this we split the data at 1 keV and considered the ratio of hard to
soft counts in (1) the two 2.5-pixel hotspot regions considered above
and (2) the remainder of the extended region, in both cases using the
same adjacent background rectangles. Although the hotspots appear
softer than the rest of the extended emission, the difference is not
significant at better than about the 1.5 sigma level, and the extended
region near to the core is more likely to be contaminated by hard
photons from the heavily absorbed core region.

Finally we considered joint fits to the {\it Chandra} and {\it
  XMM-Newton} data. {\it XMM} does not have the resolution to separate
the compact and extended regions seen with {\it Chandra} and so for
both instruments we extracted a spectrum using a 25-arcsec source circle
centered on the AGN. The spectrum of this region is obviously
dominated by the absorbed emission from the AGN itself, but will
contain a contribution from the extended component. We fitted a model
consisting of an unabsorbed power law with normalization and photon
index fixed to the values determined with the {\it Chandra}
spectroscopy, an absorbed power law with free normalization, photon
index and absorbing column, and either an APEC model with abundance
0.35 solar or a further power law. In either model the parameters of
the AGN component were well constrained and similar to those derived
from {\it Chandra} alone. For the thermal model ($\chi^2 = 168/144$)
we found $N_{\rm H} = (9.2_{-0.7}^{+1.2}) \times 10^{22}$ cm$^{-2}$ and photon
index $1.68_{-0.15}^{+0.19}$, while the thermal component had $kT =
3.1_{-1.0}^{+2.3}$ keV. The power-law model ($\chi^2 = 163/144$) had
$N_{\rm H} = (8.8_{-1.0}^{+1.3}) \times 10^{22}$ cm$^{-2}$, a photon
index of $1.64_{-0.15}^{+0.20}$, and a photon index for the additional
unabsorbed power law of $2.1 \pm 0.3$. The {\it XMM} data thus show
that the active nucleus of 3C\,171 does not seem to have varied in
normalization, photon index or absorbing column over the three months
between the two observations, but they do not really help us to
distinguish between a thermal and a non-thermal model for the extended
emission.

\section{Discussion}

\subsection{Thermal or non-thermal?}

The interpretation of these observations clearly hinges on the nature
of the central region of extended X-ray emission. As we have seen,
spectroscopy marginally favours a non-thermal origin for this emission
over an APEC plasma with abundance values typical for the intragroup
or intracluster medium. We know that the jets, hotspots and lobes of
radio galaxies can all emit non-thermal emission via the
inverse-Compton process. However, there seem to us to be several
compelling reasons for preferring a thermal model for all the extended
X-ray emission:

\begin{enumerate}
\item Non-thermal emission from jets and hotspots generally shows some
  correspondence with the radio emission; if there are offsets, the
  offsets are more typically in the direction of the nucleus;
  crucially, there are no documented offsets for hotspots where the
  hotspot counterpart lies outside the boundaries of the radio
  emission, and it is very hard to see how such a situation can arise.
  In 3C\,171 the peaks of the extended emission clearly lie outside
  the boundaries of the radio emission (Fig.\ \ref{small-scale}). The
  inner part of the extended emission is not associated with any
  feature in the knotty inner jet, whereas synchrotron emission from
  FRII jets tends to trace jet structure quite well
  \citep*[e.g.,][]{wys01,khwm05}. Inverse-Compton emission from the
  small-scale lobes would be expected to be well matched
  morphologically to the radio structure, but this is not seen in
  3C\,171: in any case the predicted inverse-Compton emission from
  3C\,171, if it follows the trend seen in other FRIIs at similar
  redshift \citep[e.g.,][]{chhb05} is negligible.

\item The good agreement between the structure in emission lines and
  the X-rays, which may extend to regions where there is no radio
  emission at all, is very hard to explain in a non-thermal model.

\item The strong correlation between X-ray emission and
  depolarization, and particularly the fact that this correlation
  appears better for the X-ray emission than for emission at any other
  waveband, is impossible to explain in a non-thermal model.
\end{enumerate}

In what follows we therefore investigate the consequences of
considering a thermal model for the extended X-ray emission in
3C\,171.

\subsection{Shock-ionized or photoionized?}

Is the extended X-ray emission thermal bremsstrahlung from
shock-heated thermal material or could it be dominated by unresolved
emission lines photoionized by the flux from the active nucleus? These
two models imply different density, temperature and pressure profiles
for the X-ray-emitting plasma. In particular, the rough constancy in
both the X-ray brightness along the jet and the X-ray to optical
emission line flux ratio on scales between $\sim 5$ and 20 kpc from
the nucleus implies a roughly constant ionization parameter if both
the optical and X-ray emission is produced by photoionization; as the
photon number density drops as $r^{-2}$, the density profile (assuming
an ionization-cone geometry) must also go roughly as $r^{-2}$. By
contrast, in a thermal bremsstrahlung model the density must be
roughly constant throughout the emitting region.

One piece of evidence favouring the photoionized model comes from the
ratio of optical emission line flux to soft X-ray flux, following
\citet*{bgc06} who analysed the extended X-ray and radio structures in
a sample of nearby Seyfert galaxies. The total flux of [OIII] emission
in a 10-arcsec aperture centered around the peak of the source is $3.9
\times 10^{-14}$ erg s$^{-1}$ cm$^{-2}$ (Tilak, priv. comm.). If we
scale this to include only the emission at distances $> 2$ arcsec from the
nucleus, it drops by a factor 2. The unabsorbed X-ray flux of the
extended emission in the 0.5-2.0 keV band used by \citeauthor{bgc06}
is $(7.5 \pm 0.7) \times 10^{-15}$ erg s$^{-1}$ cm$^{-2}$, so the
ratio of the optical to soft X-ray flux is roughly a factor 2.5. The
flux ratio seen here thus is comparable to, though at the low end of,
that seen in the Seyferts analysed by \citeauthor{bgc06}, where both
the [OIII] emission and the X-rays are inferred to be photoionized by
the nucleus. However, in 3C\,171 we know that the [OIII] emission is
most likely to be shock-ionized \citep{catr98}, so this agreement may be a
coincidence.

We can place some simple limits on the photoionized model using the
requirement of constant ionization parameter, following the analysis
of \citet{wmsw95}. Let the emitting geometry be a cone between
radial distances of $r_{\rm in}$ and $r_{\rm out}$, so that the radius
of the cone at a distance $r$ is given by $R(r) = R_{\rm out}(r/r_{\rm
  out})$, and let the density be given by $n(r) = n_{\rm in} (r/r_{\rm
  in})^{-2}$. Then the total luminosity of the cone is given by
\begin{eqnarray}
L_{\rm line}&=&\int_{r_{\rm in}}^{r_{\rm out}} \pi R^2(r) n^2(r) j(\xi) {\rm d}r\nonumber\\
&=&\pi \left({{R_{\rm out}}\over{r_{\rm out}}}\right)^2 n^2_{\rm in}
  r_{\rm in}^4 j(\xi) \left [{1\over{r_{\rm in}}} - {1\over{r_{\rm out}}}\right]
\end{eqnarray}
and we see that, if geometrical factors are fixed and the emissivity
for a given ionization parameter $\xi$ in the relevant band is known
(\citeauthor{wmsw95} give $10^{-24}$ erg cm$^{3}$ s$^{-1}$) the luminosity
determines the density. (We assume a filling factor of order unity
since we know that low filling factors will prevent the X-ray-emitting
plasma from depolarizing the radio emission; see below.) The soft
X-ray luminosity, given the flux above, is $\sim 10^{42}$ erg s$^{-1}$. If we take $r_{\rm in} = 4$
pixels $=1.97$ arcsec $= 7.4$ kpc, $r_{\rm out} = 5.5$ arcsec $= 20.7$ kpc, and
$R_{\rm out} = 5$ kpc (based on the spatially resolved optical emission-line gas, see
H03) then we find $n_{\rm in} \approx 1$ cm$^{-3}$, and so $n(r_{\rm
  out}) \approx 0.1$ cm$^{-3}$. We have the additional requirement
that the ionization parameter $\xi$, defined as
\begin{equation}
\xi = {L_{\rm s}\over nr^2}
\end{equation}
be $>100$ erg s$^{-1}$ cm \citep{wmsw95} and for the constraint
on $n$ that we have just derived this implies that the ionizing
luminosity $L_{\rm s}$ has to be $>5 \times 10^{46}$ erg s$^{-1}$;
even allowing for anisotropy of the ionizing continuum, this seems
very high compared to the X-ray luminosity of the nucleus computed in
Section \ref{spectroscopy}. Moreover, when we consider the temperature of
the photoionized gas (likely to be $\sim 10^6$ K, see \citealt{kk84}) we can see that the external pressures are likely to be
such that the radio source would drive a shock into this material
anyway (see below, Section \ref{dynamics}); so it is hard to see how a
self-consistent photoionization model can be derived. We therefore
focus in what follows on the implications of a model in which the
X-ray emission comes from collisionally ionized, shock-heated material.

\subsection{The depolarizing medium?}

Is the extended X-ray emission coming directly from the depolarizing
medium inferred from the radio data (H03)? From polarization
observations at high frequency, H03 inferred that the product $n_{\rm
  e} B d \approx 0.1$ cm$^{-3}$ nT kpc$^{-1}$, where $n_{\rm e}$ is
the electron density in the depolarizing medium, $B$ is the field
strength, and $d$ is the scale size of field reversals, required by
observations to be $\la 1$ kpc; given the
uncertainties and simplifying assumptions, this constraint is probably
only an order-of-magnitude estimate. Note in particular that H03
assumed a cylindrical geometry with radius 5 kpc and length 50 kpc for the
depolarizing medium, based on the appearance of the source in optical
emission lines.

On the assumption of a collisionally ionized, thermal origin for the
X-rays, we can use the APEC normalization to work out a mean electron
density. We begin by using the parameters of H03 for the geometry of
the depolarizing region; this gives $n_{\rm e} \approx 0.05$ cm$^{-3}$
if we assume that the emission region is a uniformly filled cylinder.
Thus, for $d<1$ kpc, we obtain a constraint on the external magnetic
field, $B > 2$ nT. H03 preferred lower field strengths, but there is
no reason in principle why they cannot be higher. A plausible {\it
  upper} limit is given by the field strength that gives an energy
density equal to that in the gas, i.e.
\begin{equation}
(3/2 \times 1.84 n_e) kT = {{B^2}\over {2\mu_0}}
\end{equation}
since we do not expect the external gas to be magnetically dominated.
This constraint gives $B < 9$ nT for $kT = 1.5$ keV, which is entirely
consistent with the lower limit derived from the depolarization
observations, and implies $d > 200$ pc for our adopted density, which
is reasonable as we know that some of the Faraday structure was
resolved by the highest-resolution observations of H03. Moreover, for
the density and temperature implied by the thermal model for the
extended X-rays, we find a pressure $p$ in the X-ray-emitting gas
equal to around $2 \times 10^{-10}$ dyn cm$^{-2}$, which is very
consistent with what might be expected for pressure balance with
emission-line clouds having $n_{\rm e} = 100$ cm$^{-3}$, kT = $10^4$
K. We conclude that it is very plausible that we are directly imaging
the depolarizing medium, in which case the depolarization data require
that the magnetic field energy density in the X-ray-emitting plasma be
within a factor of a few of the thermal energy density. A consequence
of this is that the filling factor of the X-ray emitting material must
be of order unity, since low-filling-factor plasma cannot effectively
depolarize the radio emission; thus the X-ray-based pressures should
be a good measure of the true pressure around the jet.

The choice of geometry for the hot gas affects the inferred
parameters. As mentioned above, H03's assumptions were based on the
properties of the [OIII] emission-line gas. Given the extent of the
{\it Chandra} extended emission, a length of the cylinder of around 42
kpc and a radius around 5.6 kpc (without correction for the resolution
of the instrument) might be estimated. Since $n_{\rm e}$ goes as
$V^{-1/2}$, the effect on the inferred density of adopting these
X-ray-derived parameters would be negligible; unless the source is
very far from the plane of the sky, the likely effect of any
correction for projection would be similarly limited. Accordingly we
retain the parameters used by H03 in what follows.

\subsection{Dynamics, outflows and source energetics}
\label{dynamics}

If the X-ray emitting gas genuinely represents the depolarizing
medium, and fills the space in between the cold emission-line clouds
with a filling factor close to unity, and if the emission-line clouds
acquire their observed radial velocity offsets as a result of
interaction with the jet, as implied by the association of high
velocities with jet structures seen by \citet{catr98}, then it seems very plausible that the hot
phase is moving along with the cold clouds. In particular, if the cold
clouds are swept up, shock-ionized material, as argued by Clark
\etal , then the material in between them will be swept up too.
From the line splitting observed by \citeauthor{catr98}, we can put a
lower limit on the velocity of the cold clouds away from the nucleus
of $\sim 500$ km s$^{-1}$; as mentioned above, this is highly
conservative given the evidence that the jet axis lies close to the
plane of the sky. In addition to the thermal energy stored in the gas,
$\sim (3/2) V (n_{\rm e}+n_{\rm p}) kT$, which is around $4 \times 10^{58}$ erg,
we require that the radio source must have supplied the kinetic energy
of the outflow, $\sim (1/2) n_{\rm p} m_{\rm p} V v^2$, which is $\ga 10^{58}$
erg, given that the mass of the hot gas, $n_{\rm p} m_{\rm p} V$, is around $3
\times 10^9 M_\odot$. However, if the outflow is driven by a shock,
the expansion speed $v$ may be much higher. We next consider the
dynamics and energetics of a possible shock in 3C\,171.

We may compare the pressures derived in the previous section to the
minimum energy in the radio components in order to gain some insight
into the source dynamics. In the brightest region of the E hotspot,
which we model as a cylinder with a length of 0.7 arcsec and a width
of 0.225 arcsec based on the high-resolution radio images of H03, we
fitted a broken power-law model in electron energy to the electron
energy spectrum to determine the minimum energy. We conservatively
assume no protons and $\gamma_{\rm min} = 1000$, and find a minimum
pressure (assuming a fully tangled field, so that $p_{\rm min} =
U_{\rm min}/3$) of around $4 \times 10^{-8}$ dyn cm$^{-2}$. This is
substantially higher than the external pressure inferred in the
previous section, so it seems clear that the hotspot and the region of
the jet head around it are capable of driving a shock into the
external medium, explaining the presence both of the hot and cold
phases of ionized gas. Modelling this bright region of the hotspot as a
uniform cylinder significantly understates the complexity of the radio
emission, but no plausible more detailed model would reduce the
minimum pressure by 1.5 orders of magnitude. On the other hand, the
minimum pressure in the large-scale lobes, on scales of tens of kpc,
is $\sim 3 \times 10^{-11}$ dyn cm$^{-2}$, giving a plausible reason
why these large-scale lobes avoid the central regions of the source.

If we assume that the minimum pressure in the lobes, estimated above,
is comparable to (it clearly cannot be less than) the pressure in any
external hot-gas environment, then we can work out the Mach number of
the shock being driven by the jets, given that
\begin{equation}
{{p_2}\over p_1} = {{2\Gamma{\cal M}_1^2 + (1-\Gamma)}\over{\Gamma +1}}
\end{equation}
where $\Gamma$ is the adiabatic index, 5/3 in this case; this implies
a moderate-strength shock of ${\cal M}_1 \sim 2.4$, a density contrast
across the shock of around 2.6, and a pre-shock temperature of 0.6
keV, in turn implying a shock speed around 1400 km s$^{-1}$. These
values seem plausible. The temperature estimate for the hot gas is
consistent with what might be expected for a poor group of galaxies.
From the temperature-luminosity relationship for groups
\citep[e.g.,][]{op04}, we expect a total bolometric luminosity for
such a system on the order of $4 \times 10^{41}$ erg s$^{-1}$, which
is comparable to the upper limit on the group-scale bolometric
unabsorbed luminosity for $kT = 0.6$ keV if we assume there are $<30$
counts in group-scale emission, $< 7 \times 10^{41}$ erg s$^{-1}$.
Finally, the required external density on scales of a few kpc, $n_{\rm
  e} \sim 0.02$ cm$^{-3}$, although a little high, seems to be within
a factor of a few of what is plausible for a poor group (see, e.g.,
the entropy profiles of \citealt*{psf03}). Thus all the evidence
seems consistent with the idea that the X-ray emission is shocked gas
from a powerful jet propagating into a poor group environment, though
we cannot rule out the possibility that the shock is instead being
driven through much colder, denser external material which would not
be visible in the X-ray (a point we return to below, Section \ref{otherss}).
If we assume that the hot and cold phases are moving out with the jet
at 1400 km s$^{-1}$, then the kinetic energy of the outflow becomes
$\sim 8 \times 10^{58}$ erg.

If the estimates of shock speed above are correct, then the projected
jet length of around 16 kpc implies a source lifetime $\sim L/v
\approx 10^7$ years, which is not atypical for dynamic age estimates
of small radio galaxies (projection would imply longer timescales).
The jet then has to supply at least the excess thermal and kinetic
energy of the gas, which is $\sim 10^{59}$ erg for $v \sim 1400$ km
s$^{-1}$, on this timescale, giving a required total jet power of $3
\times 10^{44}$ erg s$^{-1}$. This is in one sense a lower limit on
the jet power since it excludes the energy stored in the lobes, but in
another sense an upper limit since it assumes that all the gas moves
at the putative shock speed and neglects any projection effects.
Nevertheless it is interesting that this jet power is very comparable
to the bolometric radiative AGN power that can be estimated from the
unabsorbed X-ray luminosity (Section \ref{spectroscopy}), given that widely
used bolometric corrections from the 2-10 keV band are an order of
magnitude or more \citep[e.g.,][]{mrgh04}.

\subsection{3C\,171, 3C\,305, and other radio galaxies}
\label{otherss}

So far we have not addressed the question of the difference between
3C\,171 and most other radio galaxies of similar radio luminosity,
which have classical double radio morphology with double lobes that
extend back from the hotspots to form a `cocoon' around the radio
jets. It is hard to escape the impression that the peculiar radio
morphology of 3C\,171 is a {\it result} of its unusual extended
emission-line region and the associated X-ray-emitting and
radio-depolarizing medium. This impression is strengthened by the
similarity between 3C\,171 and 3C\,305. 3C\,305 is a lower-power,
nearer, smaller radio source, but also has twin FRII-like jets and
hotspots \citep{hmbv82,jbpc03} together with
disrupted, perpendicular plumes rather than conventional lobes
(\citealt{hmbv82}), and has jets associated with optical emission-line
gas that is itself associated with radio depolarization \citep{hmbv82}, that is likely shock-ionized \citep{jbpc03} and
is spatially coincident with extended X-ray emission that has been
imaged with {\it Chandra} \citep{mcgg09}. In 3C\,305, an
outflow of neutral hydrogen is also associated with the jet \citep{motv05}. These two radio galaxies, 3C\,171 and 3C\,305, seem to
us to be the best-studied examples of a class of powerful radio-loud
AGN that is relatively rare, at least at low redshift, in which there
is a strong, disruptive interaction between the IGM of the host galaxy
and the powerful, FRII-type radio jets. (We will discuss the
interpretation of the radio depolarization in 3C\,305 and its
relationship to the extended X-ray emission in a future paper.)

What property of the IGM makes 3C\,171 and 3C\,305 special? Although,
for convenience, we have discussed the dynamics of the source in terms
of propagation into the hot, high-pressure phase of the IGM, it seems
clear that special properties of this phase cannot account for the
peculiar structures of these two radio sources. All powerful FRII
sources seem to propagate into environments of at least comparable
richness, and many inhabit richer groups \citep[e.g.,][]{cbhw04}
while detailed observations have shown that quite strong asymmetries
in the distribution of intragroup hot gas \citep[e.g.,][]{hkwc07} do
not produce radio sources as distorted as 3C\,171 and 3C\,305. We are
therefore forced to the conclusion that the strong interaction is
between the radio source and the {\it cold} phase of the IGM. In many
radio galaxies the amount of cold gas is negligible, but we know that
it is significant at least in 3C\,305 \citep{motv05}. We propose that
both radio sources have acquired their peculiar structure by
propagating into initially cold material that is dense enough to be
dynamically important on kpc scales. The shock-heating of this
material produces both the X-ray emission and the optical emission
lines that we see. Even after the shock, the material along the jet
axis is dense enough that the thermal pressure forces the post-hotspot
outflow in a direction transverse to the jet, causing it to expand
into the observed perpendicular plume structures. This picture
naturally explains why these sources are rare at the present day --
large masses of cold gas aligned with the jet on kpc scales are not
usually present in the elliptical hosts of powerful radio galaxies.
However, we expect them to be much more common in the early universe;
an example may be provided by the morphologically very similar, though
much larger, X-ray structures seen along the jet axis of the powerful
$z=2.2$ radio galaxy PKS 1138$-$262 \citep{chpr02}. If this is
the case, their jet-driven (rather than radiatively driven) heating
and expulsion of cold gas from the central regions of the host galaxy
may be related to observational constraints on the masses and
star-formation histories of massive elliptical galaxies
\citep[e.g.,][]{t07} and their study at low redshift gives us
important insights into the process of high-redshift massive galaxy
formation.

\section{Summary and conclusions}

Our main results on 3C\,171 can be summarized as follows.

\begin{enumerate}
\item We find an extended region of X-ray emission coincident with the
  well-studied extended optical emission-line region in 3C\,171.
\item In what we feel is the most plausible interpretation, the X-ray
  emission is thermal and is due to collisionally ionized hot plasma
  in which the cold clouds of optical line-emitting material are
  embedded. On these scales photoionization models require very high
  continuum luminosities and are hard to render consistent with the
  dynamics of the source and the observed radio depolarization.
\item In the collisionally ionized interpretation, the X-ray emitting
  material is capable of producing the depolarization of the radio
  emission from the jets and hotspots studied by H03. The physical
  conditions we estimate from the X-ray observation give rise to very
  reasonable constraints on the magnetic field strength in the medium
  external to the radio source, $2\ \mathrm{nT} < B < 9\ \mathrm{nT}$,
  and a scale for the magnetic field reversals in the range
  $0.2\ \mathrm{kpc} < d < 1.0\ \mathrm{kpc}$.
\item If the hot plasma is moving with the cold line-emitting
  material, then it dominates the energetics of the outflow. The total
  energy budget required depends on the outflow speed adopted, but
  lies in a range between a few $\times 10^{43}$ and a few $\times
  10^{44}$ erg s$^{-1}$, comparable to the bolometric radiative
  luminosity of the active nucleus.
\end{enumerate}

These results are significant for two reasons. Firstly, the very good
agreement between the physical conditions implied by the optical line
emission, the X-rays and the radio depolarization gives us some
confidence that we have at least a basic picture of the complex
multi-phase medium (consisting of at least warm gas, hot plasma and
magnetic fields) in which the jet is embedded. Over the coming
decades, our ability to study polarization structures in radio
galaxies will be greatly improved, while our ability to image them
with high resolution in X-rays may unfortunately be significantly
diminished. It is therefore important to develop methods of inferring
physical conditions from radio polarization (and optical
emission-line) studies alone, and our work shows that this can be
done. Secondly, they imply that many of the sources in which extended
emission-line regions have been detected may also have energetically
dominant outflows of hot plasma; if confirmed by X-ray studies of
suitable targets, this makes optical emission-line studies central to
an understanding of the interaction between radio galaxies and the
cold gas in their environments.

\section*{acknowledgements}

We thank Judith Croston for helpful discussions of the central density
in poor groups, Clive Tadhunter for providing emission-line images,
and Avanti Tilak for help with the [OIII] luminosity of 3C\,171. MJH thanks the
Royal Society for a research fellowship. FM acknowledges the
Foundation BLANCEFLOR Boncompagni-Ludovisi, n\'ee Bildt, for the grant
awarded him in 2009 to support his research. Work at SAO was supported
by NASA grant GO9-0110X.

\end{document}